\documentclass{webofc}
\usepackage[varg]{txfonts}   

\usepackage[utf8]{inputenc}
\usepackage{graphicx}
\usepackage{dcolumn}
\usepackage{bm}
\usepackage{feynmf}
\usepackage{color}

\newcommand{\be}{\begin{equation*}}
\newcommand{\ee}{\end{equation*}}
\newcommand{\ba}{\begin{eqnarray*}}
\newcommand{\ea}{\end{eqnarray*}}
\newcommand{\ban}{\begin{eqnarray}}
\newcommand{\ean}{\end{eqnarray}}

\newcommand{\bw}{\begin{widetext}}
\newcommand{\ew}{\end{widetext}}

\newcommand{\pppp}{\ensuremath{\mathord{+}\mathord{+}\mathord{+}\mathord{+}}}
\newcommand{\mmmm}{\ensuremath{\mathord{-}\mathord{-}\mathord{-}\mathord{-}}}
\newcommand{\pmpm}{\ensuremath{\mathord{+}\mathord{-}\mathord{+}\mathord{-}}}

\newcommand{\Imag}{\mathop{\mathrm{Im}}}

\begin{document}

\title{Unitarized one-loop graviton-graviton scattering}

\author{\firstname{Rafael} \lastname{L. Delgado}\inst{1}\fnsep\thanks{\email{rafael.delgado@upm.es}}
        \firstname{Antonio} \lastname{Dobado}\inst{2}\fnsep\thanks{\email{dobado@fis.ucm.es}}
        \firstname{Domènec} \lastname{Espriu}\inst{3}\fnsep\thanks{\email{espriu@icc.ub.edu}}
        }

\institute{Matemática Aplicada a las TIC, ETSIS de Telecomunicación,
  Universidad Politécnica de Madrid (Campus Sur) 28031 Madrid, Spain
    \and
  Departamento de Física Teórica and IPARCOS,
  Universidad Complutense de Madrid, 28040 Madrid, Spain
    \and
  Departament de Física Quàntica i Astrofísica and Institut de Ciències del Cosmos (ICCUB),
  Universitat de Barcelona, 08028 Barcelona, Catalonia, Spain
}

\abstract{

In this work we interpret the Einstein-Hilbert (EH) Lagrangian of gravitation as the first term of a low-energy effective theory similar to those considered in the chiral Lagrangian approach to low-energy hadron physics or the electroweak chiral Lagragians describing the symmetry breaking sector of the Standard Model (SM). Starting from the one-loop computation of the elastic graviton-graviton scattering amplitude by Dunbar and Norridge, we unitarize the IR regularized partial waves by using the Inverse Amplitude Method (IAM). This method enlarges the regime of applicability of the perturbative results to higher energies of the order of the Plank scale $M_P$ and 
allows for the possibility of poles in the second Riemann 
which have the natural interpretation of dynamical
resonances. In this work we look for these possible resonances for the $\pppp$ and $\mmmm$ helicity channels and the $J=0$, $2$ and $4$ partial waves.}
\maketitle

\section{Introduction}
As it is well known the Einstein-Hilbert (EH) Lagrangian can be considered as the first
term of an effective theory for gravity~\cite{Donoghue:1994dn,Donoghue:1995cz,Donoghue:2019jeq,Espriu:2009ju}, being the expansion parameter the typical energy of the process considered over the Plank mass scale $M_P$. The situation is similar to that of the Chiral Perturbation Theory 
describing low-energy pion dynamics where the role of $M_P$ is played by the pion decay constant $f_{\pi}$. Concerning graviton elastic scattering the EH term produces an amplitude which is of the order $s/M_P^2$ and was first computed by~\cite{DeWitt:1967uc,Berends:1974gk,Grisaru:1975bx}. The one-loop  corrections 
were obtained by Dunbar and Norridge by means of string theory methods~\cite{Dunbar:1994bn}. In principle one would expect UV divergences but as far as the only possible counterterms are proportional to $R^2$, $R_{\mu\nu}\,R^{\mu\nu}$ or  $R^{\;\;\;\;\gamma\delta}_{\alpha\beta}\,R^{\;\;\;\;\alpha\beta}_{\gamma\delta}$, and these counterterms vanish on-shell, the one-loop result is UV finite as it was stated by Hooft and Veltman~\cite{tHV}. However, due to the fact that gravitons are massless, IR divergences appear.

In~\cite{Delgado:2022qnh} the authors considered the possibility of extending the one-loop amplitude to higher energies by using the IAM unitarization procedure. By defining the initial momenta and helicities by $p_1$, $\lambda_1$, $p_2$, $\lambda_2$,
and the final ones by $p_3$, $\lambda_3$, $p_4$ and $\lambda_4$, the helicity amplitude can be written as:
\begin{equation}
T_{\lambda_1\lambda_2\lambda_3\lambda_4}(s,t,u)=\left<p_3,\lambda_3;p_4,\lambda_4\mid T \mid p_1,\lambda_1;p_2,\lambda_2 \right>,
\end{equation}
where the usual Mandelstam variables $s=(p_1+p_2)^2, t=(p_1-p_3)^2$ and $u=(p_1-p_4)^2$ are used and $T$ is the standard reaction matrix $S= I + i(2\pi)^4\delta^{(4)}(P_f-P_i)T$,
with $P_i=p_1+p_2$, $P_f= p_3+p_4$ and $S$ being the $S$-matrix. The helicities $\lambda_i$ take the values $+2$ and $-2$, denoted by $\lambda_i=+$ and $-$ respectively.
By using $P$ and $T$ invariance and crossing, all the amplitudes can be written in terms of
\begin{equation}
A(s,t,u)=T_{++++}(s,t,u),\quad
B(s,t,u)=T_{+++-}(s,t,u),\quad
C(s,t,u)=T_{++--}(s,t,u).  
\end{equation}
At tree level,
\begin{equation}
A^{(0)}(s,t,u) = \frac{8\pi}{M_P^2}\frac{s^3}{tu},\quad
B^{(0)}(s,t,u) = 0,\quad
C^{(0)}(s,t,u) = 0.
\end{equation}
The NLO amplitudes $A^{(1)}(s,t,u)$, $B^{(1)}(s,t,u)$ and $C^{(1)}(s,t,u)$, obtained by Dunbar and Norridge~\cite{Dunbar:1994bn}, can also be found in~\cite{Delgado:2022qnh}.

\section{Unitarization}
In order to perform the unitarization of the amplitude we need to compute the partial waves which in principle are defined by
\begin{equation}
a_{J\lambda_1,\lambda_2,\lambda_3\lambda_4}(s)=\frac{1}{64\pi}\int_{-1}^{1}d(\cos \theta)   d^J_{\lambda,\lambda'}(\theta) T_{\lambda_1\lambda_2\lambda_3\lambda_4} (s,\theta),
\end{equation}
where $\lambda=\lambda_1 - \lambda_2$, $\lambda'=\lambda_3 - \lambda_4$, $t=-(s/2)(1-x)$, $u=-(s/2)(1+x)$ and $x=\cos\theta$.

However, in the case of graviton-graviton scattering, these integrals are not well defined, even at the tree level, since the helicity amplitudes diverge for $x=\cos\theta$ close to $\pm 1$. Hence, we regularize~\cite{Delgado:2022qnh} the helicity amplitudes as
\begin{equation}
\tilde T^\eta_{\lambda_1\lambda_2\lambda_3\lambda_4} (s,\theta) = T_{\lambda_1\lambda_2\lambda_3\lambda_4} (s,\theta)
\end{equation}
if and only if $\cos \theta\in (-1+\eta,1-\eta)$, and $\tilde T^\eta_{\lambda_1\lambda_2\lambda_3\lambda_4}=0$ otherwise.
With this definition, the re-defined partial waves are:
\begin{equation}
a_{J\lambda_1,\lambda_2,\lambda_3\lambda_4}(s,\eta)=
\frac{1}{64\pi}\int_{-1+\eta}^{1-\eta}d(\cos \theta) d^J_{\lambda,\lambda'}(\theta) T_{\lambda_1\lambda_2\lambda_3\lambda_4} (s,\theta),
\end{equation}
and the expected elastic scattering unitarity relation would be:
\begin{equation}
\Imag \tilde T_{\lambda_1\lambda_2\lambda_3\lambda_4}^\eta (s,\theta) = 
\frac{1}{128 \pi^2}\sum_{\lambda_a \lambda_b}
\int_R d \Omega' 
\tilde T_{\lambda_1\lambda_2\lambda_a\lambda_b}^\eta (s,\theta')
\tilde T_{\lambda_a\lambda_b\lambda_3\lambda_4}^{\eta*} (s,\theta''),
\end{equation}
where $R$ is the region of the two-body phase-space defined by $s=4E_{\rm CM}^2 > \mu^2$, with $\mu$ being an IR regulator.
The key point of our work~\cite{Delgado:2022qnh} is that both $\eta$ and $\mu$ are related. This is because crossing requires $t<-\mu^2$ and by trading $\cos\theta$ by $t$:
\begin{equation}
    \int^{1-\eta}_{-1+\eta}d(\cos\theta)= \frac{s}{2}\int^{t_{max}}_{t_{min}}dt=\frac{s}{2}\int^{-\mu^2}_{-s+\mu^2}dt,
\end{equation}
leading to
\begin{equation}\label{mu_eta_rel}
    \mu^2=\eta\frac{s}{2}.
\end{equation}
By using this relation, one finds for the $\pppp$ and $\mmmm$ partial waves:
\begin{equation}
    \Imag a^{(1)}_0 (s,\mu) = \left(a^{(0)}_0 (s,\mu)\right)^2
\end{equation}
to all orders in $\eta$, even if we should expect such expression to hold only for small values of $\eta$. For $J\neq 0$, this
works only up to order $\eta$,
\begin{equation}
    \Imag a^{(1)}_J = \left(a^{(0)}_J\right)^2 + O(\eta).
\end{equation}
Finally, for the $\pmpm$ case, the equation is fulfilled up to constant terms, only applying for the divergent terms as $\eta\to 0$.

In the following we will focus on the $\pppp$ and $\mmmm$ partial waves with $\eta=2\mu^2/s$. The NLO partial wave is proportional to $s^2$ and has different contributions with 
$\log (s/\mu^2)$, $\log^2 (s/\mu^2)$ and $F(s/\mu^2) = \log (s/\mu^2) - \log (-s/\mu^2)$ factors~\cite{Delgado:2022qnh}. Hence, our NLO amplitudes have both right (RC) and left (LC) cuts on the complex plane and, as expected, they break (elastic) unitarity at high energies $s\simeq M_P^2$. Thus we consider the use of an unitarization procedure to regularize this behaviour, as it was done in low-energy hadronic physics \cite{Truong:1988zp,Dobado:1989qm,Dobado:1992ha,Dobado:1996ps}. Unitarization techniques can also be applied to the elastic scattering of the longitudinal components of the electroweak bosons. In~\cite{Delgado:2015kxa} different unitarization procedures were compared. Here we use the Inverse Amplitude Method (IAM)~\cite{Truong:1988zp,Dobado:1989qm,Dobado:1992ha,Dobado:1996ps,Dobado:1989gr}, since it looks more appropriate for the one-loop elastic graviton scattering amplitudes where one starts from an expansion in powers of $s/M_P^2$, with perturbative elastic unitarity and good analytical properties (RC and LC cuts). 

Thus the IAM partial waves for $\pppp$ and $\mmmm$ helicity partial waves are defined as
\begin{equation}\label{eq_IAM}
	a^{\rm IAM}_J (s,\mu) = a^{(0)}_J (s,\mu) \frac{a^{(0)}_J(s,\mu)}{a^{(0)}_J(s,\mu)-a^{(1)}_J(s,\mu)}.
\end{equation}
One of the interesting properties of these unitarized partial waves is that they can show poles on the second Riemann sheet of the complex plane under the RC that, according to
general S-matrix theory~\cite{Eden:1966dnq,Gribov:2022zpy,Novozhilov:1975yt}, would have an interpretation of dynamical resonances
due to graviton-graviton interaction. The resonance mass $M_R$ and width $\Gamma_R$ associated to a pole on position $s_0$ would be
\begin{equation}
	s_0 = M_R^2 - iM_R\Gamma_R.
\end{equation}
However, if the poles appear on the first Riemann sheet, they are most probably artifacts of the unitarization procedure (ghosts).

\section{Results}

\begin{figure*}[p]
    \centering
    \includegraphics[width=.8\textwidth]{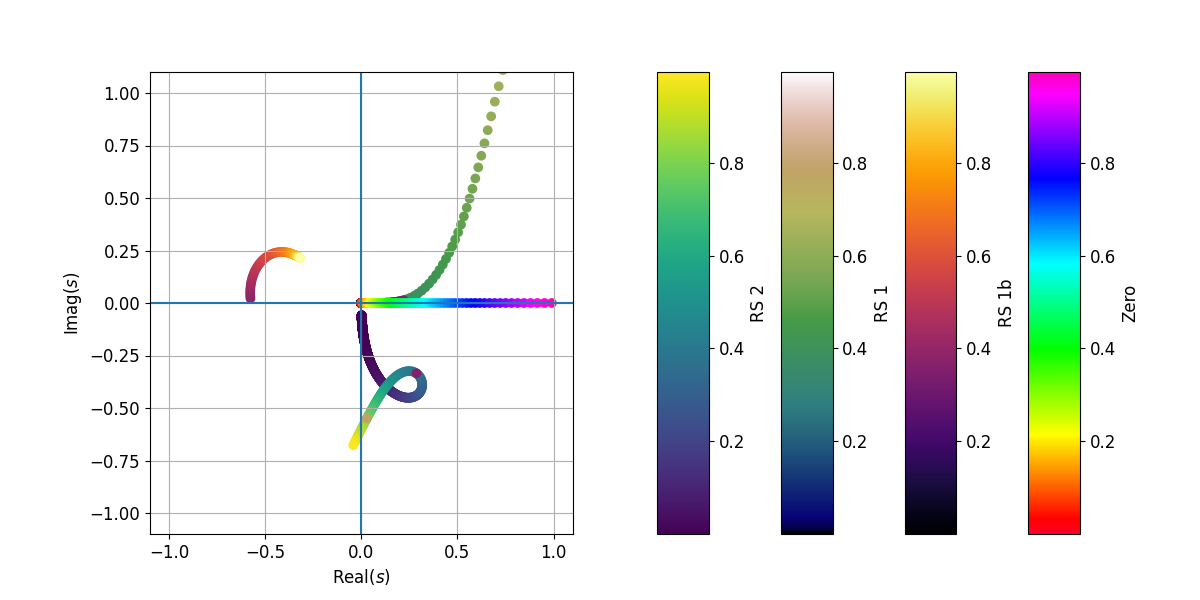}\\
    \caption{Pole positions of $J=0$ partial wave. The color scales stand for $\mu/M_P$. The \texttt{Zero} points refer to a zero in the numerator of the Inverse Amplitude Method (mostly over the positive real axis); \texttt{RS~1}, \texttt{RS~1b} and \texttt{RS~2}, to poles on the quadrants I, II and IV; and \texttt{Zero}, to zeros on the quadrant I. Figure from our ref.~\cite{Delgado:2022qnh}.}
    \label{fig:poleposition}
\end{figure*}

\begin{figure*}
    \centering
    \includegraphics[width=.45\textwidth]{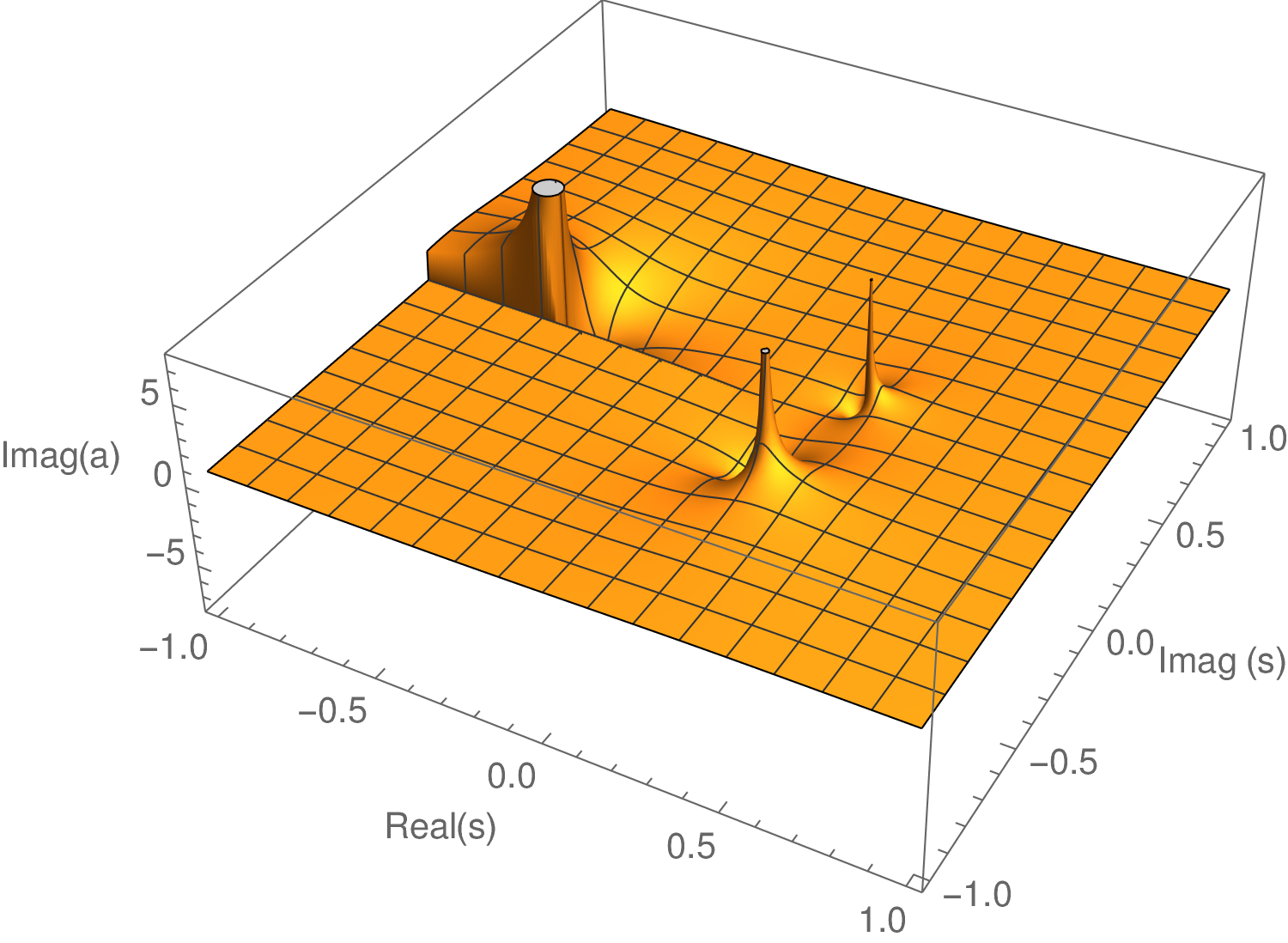} \hfill
    \includegraphics[width=.45\textwidth]{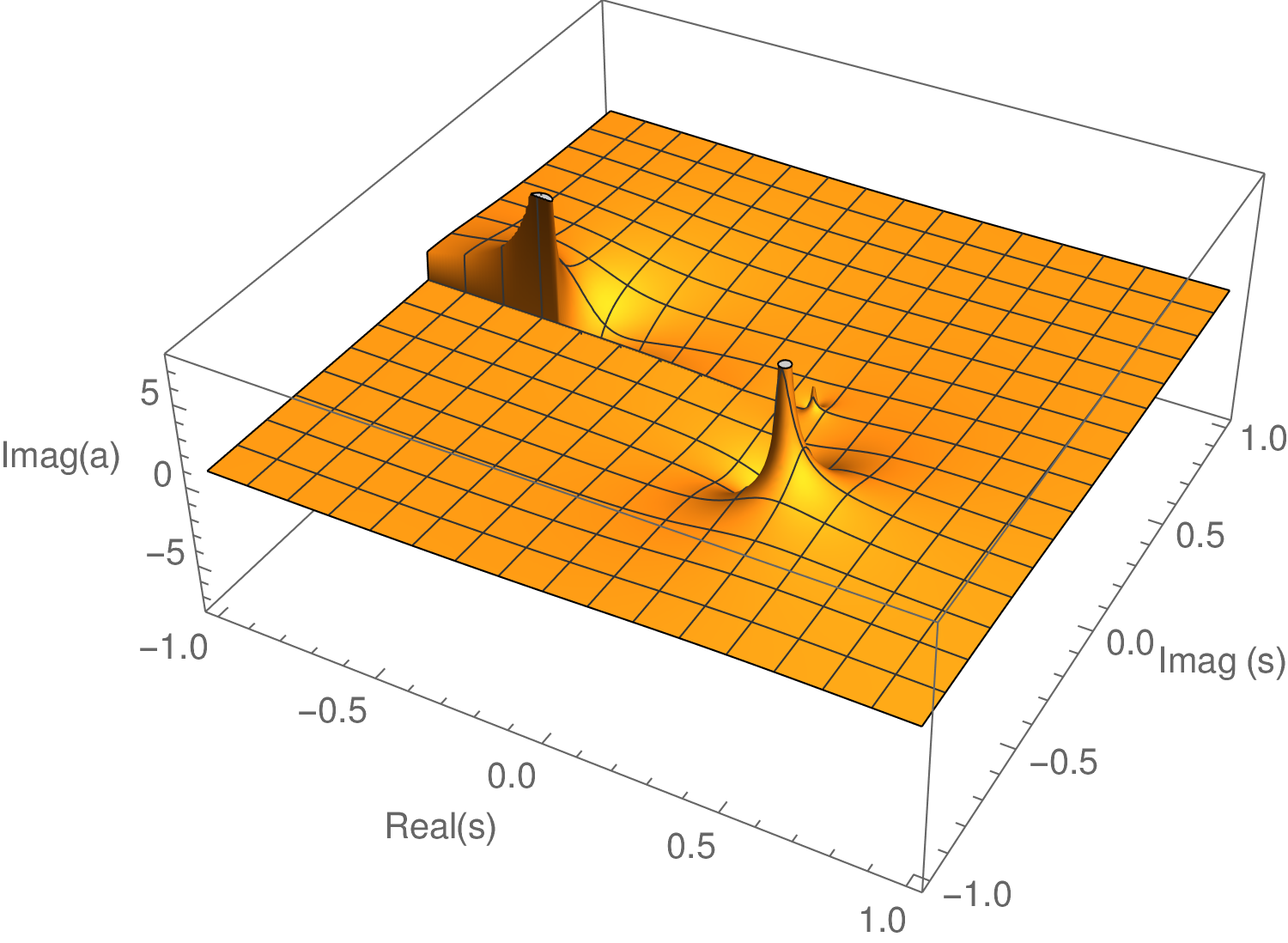} \\
    \includegraphics[width=.45\textwidth]{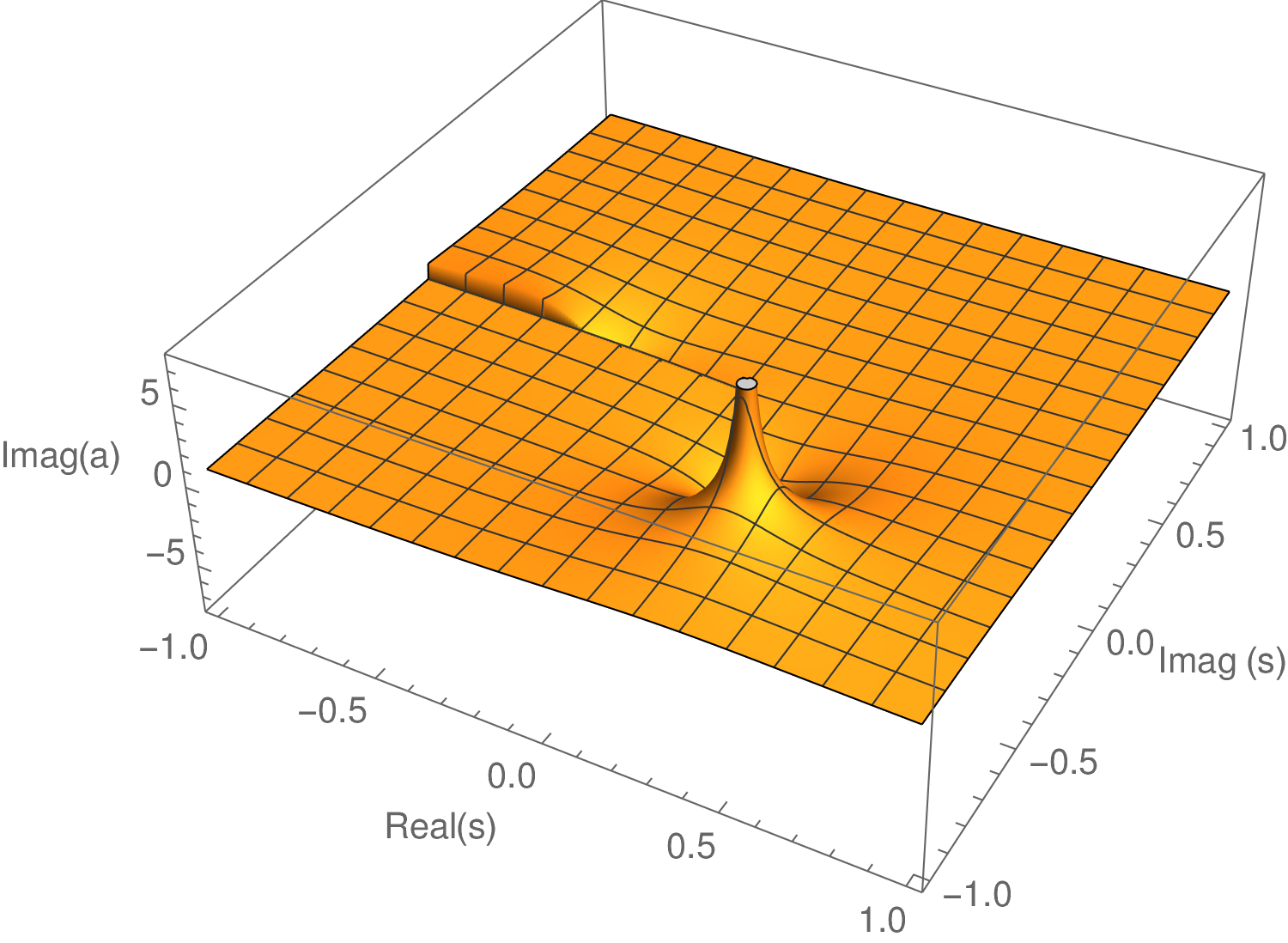} \hfill
    \includegraphics[width=.45\textwidth]{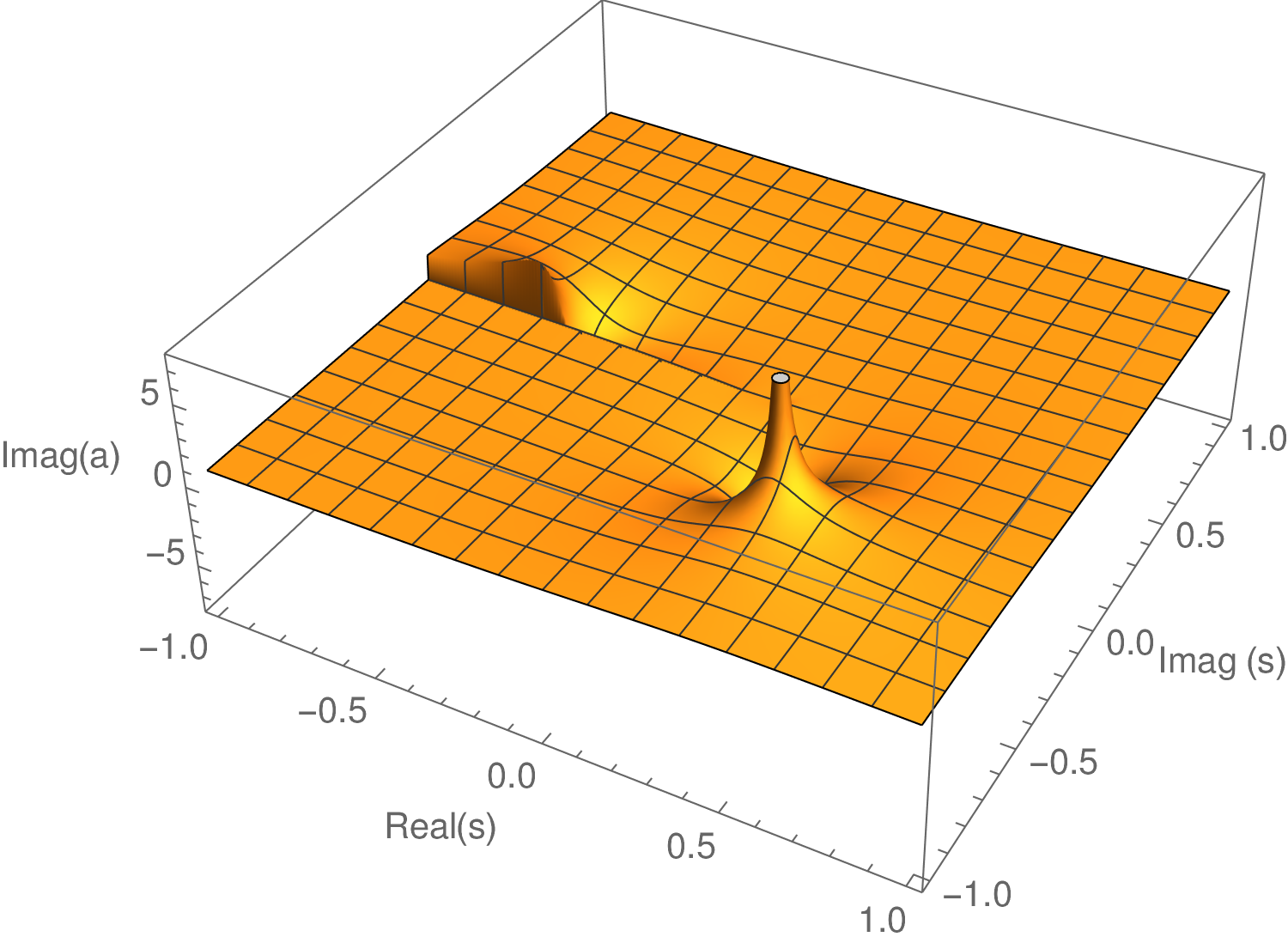} \\
    \caption{From top left, clockwise: plots of $\Imag a_0^{IAM}$ for $\mu/M_P=0.40,\,0.35,\,0.30,\,0.20$. Notice the disappearance of the poles on the first Riemann sheet (quadrants I and II). Figure from our ref.~\cite{Delgado:2022qnh}.}
    \label{fig:plots_J0}
\end{figure*}

Our preliminary results were presented in detail in~\cite{Delgado:2022qnh} where we looked for all the poles in the different Riemann sheets for values of the IR regulator $\mu$ ranging from $M_P$ to $10^{-8}M_P$. Due to the analytical structure of the perturbative NLO amplitudes and the IAM, the first and second Riemann sheets
match over the first quadrant and in particular on the positive real axis when it is approached from above. The first Riemann sheet below the real axis (negative values of $\Imag s$) is a 
mirror reflection of the amplitude above the real axis due to
Schwartz reflection principle~\cite{Eden:1966dnq,Gribov:2022zpy,Novozhilov:1975yt,Delgado:2022qnh}. Hence, since the second Riemann sheet is the analytical continuation of the amplitude across the RC,
we can find all the poles, both in first and second Riemann sheets, by analytical continuation of
the perturbative partial waves from the first quadrant.

There are two ghosts, one pole in the first and another in the second quadrant. The one in the second (\verb+RS 1b+ on fig.~\ref{fig:poleposition}) disappears for $\mu/M_P<0.37$ in the $J=0$ channel and for $\mu/M_P<0.092$ in all the channels $J=0,\,2,\,4$. The pole in the first quadrant (\verb+RS 1+ on fig.~\ref{fig:poleposition}) is a bit trickier. In this case there is a zero in the numerator of the IAM formula (eq.~\ref{eq_IAM}) and a zero in the denominator. As it can be seen on fig.~\ref{fig:poleposition} for $J=0$ (look the zero position, \verb+Zero+ on the legend) these two zeros converge at the same point as $\mu$ goes to zero. In any case, for $\mu/M_P<0.05$, the pole on the first Riemann sheet disappears for all values $J=0,\,2,\,4$  for the considered numerical resolution (see in fig.~\ref{fig:plots_J0} the $J=0$ case). 

There is also a pole on the second Riemann sheet for low values of $\mu/M_P<0.05$, which could in principle be considered as a dynamical resonance. However, the pole position shows a logarithmic dependence on $\mu/M_P$ such that it tends to the origin in the $\mu/M_P\to 0^+$ limit. This happens for the $J=0,2,4$ channels for different $\mu/M_P$ range of values.

\section{Discussion}

\begin{figure*}
    \centering
    \null\hfill
    \includegraphics[width=.65\textwidth]{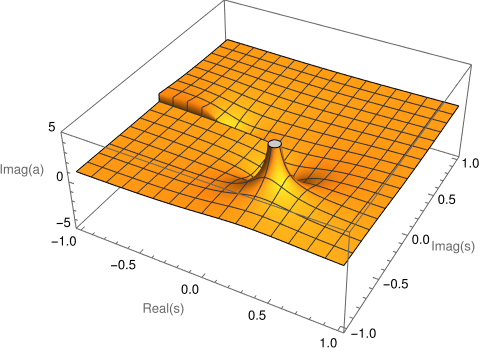} \hfill\null\\
    \includegraphics[width=.45\textwidth]{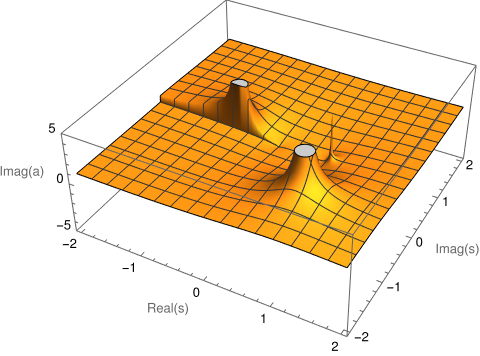} \hfill
    \includegraphics[width=.45\textwidth]{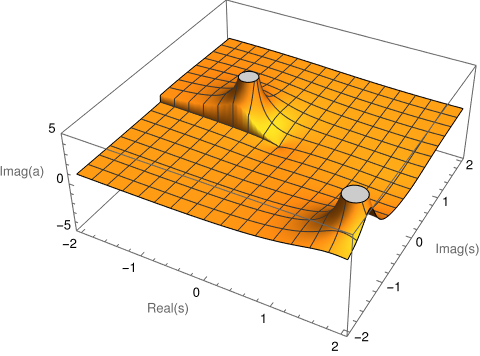} \hfill\null\\
    \caption{From top, anticlockwise: plots of $\Imag a_0^{IAM}$ for $J=0$, $J=2$ and $J=4$; $\mu/M_P=
    0.176$.}
    \label{fig:plots_Oller}
\end{figure*}
By using the IAM unitarized one-loop amplitudes for elastic graviton scattering (which depends on the IR regulator $\mu$) we have obtained (for $\pppp$ and $\mmmm$ helicities) $J=0,2,4$ partial waves having the right analytical and unitarity properties. Then we have studied the different poles of these partial waves. We found that for low enough $\mu/M_P$ values the ghosts (poles in the first and second quadrant) tend to disappear.

We also found a pole in each channel on the fourth quadrant which potentially could be understood as dynamical resonances. However these poles tend (logarithmically) to the origin of the complex plane as $\mu/M_P$ goes to zero. Therefore the physical interpretation of these poles is doubtful, to say the least, since $\mu$ was introduced as an IR regulator.

Starting from the tree level amplitude the authors of~\cite{Blas:2020och,Blas:2020dyg,Oller:2022tmo,Oller:2022ozd} found a pole in the $J=0$ channel at
$s_0=\pi x M_p^2\approx (0.22-i0.63)M_P^2$ with $x=0.07-i0.2$ which they claim is a pure gravitational resonance (graviball). This particular $x$ value corresponds to some arbitrary but reasonable choice of an UV cut-off $\Lambda$ which is needed for their unitarization method (in fact the position of the pole depends logarithmically on $\Lambda$). However notice that the width associated with this pole would be so large compared with its mass that hardly could it be considered a physical state in the usual sense. 

In any case, it seems to be interesting to check if it is possible to reproduce this pole by using the IAM unitarized NLO computation considered in this work. Thus we have looked for the closest pole we can find in our computations by minimizing the distant to their $s_0$ on the complex $s$ plane by varying our $\mu$ IR regulator. Thus we have found a pole at $s'_0=(0.23-i0.45)M_P^2$ corresponding to $\mu/M_P=0.176$. However, for this $\mu$ value ghosts are present in the $J=2$ and $J=4$ channels (see fig.~\ref{fig:plots_Oller}). Also this $\mu$ value is too high to be considered an IR cut-off at all. Further investigations are required to completely clarify this point.

\section{Acknowledgements}
We thank J.A. Oller for useful discussions and comments. This research is partly supported by the Ministerio de Ciencia e Innovación under research grants PID2019-108655GB-I00/AEI/10.13039/501100011033, PID2019-105614GB-C21 and the ``Unit of Excellence María de Maeztu 2020-2023'' award to the Institute of Cosmos Sciences (CEX2019-000918-M), and the grant 2017-SGR-929 from Generalitat de Catalunya. R. L. Delgado was also financially supported by the Ramón Areces Foundation post-doctoral fellowship, the Istituto Nazionale di Fisica Nucleare (INFN) post-doctoral fellowship AAOODGF-2019-0000329 and the Ministerio de Ciencia e Innovación PID2021-124473NB-I00.

\bibliography{references}

\end{document}